\documentclass[preprint2,10pt]{aastex}

\slugcomment{To Appear in the Astrophysical Journal}

\shorttitle{A Soft X-ray Component in the Abell 754 Cluster}
\shortauthors{Henriksen & Hudson}

\begin{document}

\title{A Soft X-ray Component in the Abell 754 Cluster}

\author{Mark J. Henriksen\altaffilmark{1}, Daniel S. Hudson, and Eric Tittley}
\affil{Joint Center for Astrophysics, Physics Department, University of Maryland,
    Baltimore, MD 21250}

\altaffiltext{1}{Laboratory for High Energy Astrophysics, NASA/GSFC}

\begin{abstract}

We have analyzed the Chandra, {\it BeppoSax}, and {\it ROSAT} PSPC observations 
of Abell 754 and report evidence of a soft, diffuse X-ray component. 
The emission is peaked in the cluster center and is detected out to 8 arcmin from the
X-ray center. 
Fitting a thermal model to the combined {\it BeppoSax} and PSPC spectra 
show excess emission below 1 keV in the PSPC and above 100 keV in the {\it BeppoSax} PDS. The source, 26W20 is
in the field of view of the PDS.
The addition of a powerlaw with the spectral parameters measured by Silverman et al. (1998) for 26W20
successfully models the hard component in the PDS.
The remaining excess soft emission can be modeled by either a
low temperature, 0.75 - 1.03 keV component, or by a powerlaw with a steep spectral index, 2.3.
Both additional components provide a statistically significant improvement over a single thermal component. Adding a non-thermal
component is an
improvement over the single thermal model with 99.7\% confidence. However, addition of a second
thermal component model provides a much better fit to the data than does the addition of a non-thermal component.
The Chandra temperature map provides a detailed description of the thermal state of the
gas on a scale of 100 kpc and larger and does not show any region cooler than 6.9 keV (90\% confidence)
within the region where the cool component was detected. Simulations of the emission from embedded 
groups are performed and compared with the Chandra temperature map which 
show groups are a plausible source of $\sim$1 keV emission. 
The cool component is centrally peaked in the cluster and the gas density and temperature
are relatively high
 arguing against the WHIM as the source of the X-ray emission. 
The typical X-ray emission from elliptical galaxies is not high enough to provide
the total cool component luminosity, 7.0$\times$10$^{43}$) ergs s$^{-1}$. 
The peak of the cool component is located between the low frequency radio halos
thus arguing against a non-thermal interpretation for the emission based on the synchrotron inverse-Compton
model which requires that the non-thermal X-ray and radio emission be co-spatial.
Thus, we conclude that emission from embedded groups is the most likely origin of the cool component
in Abell 754.



\end{abstract}


\section{Introduction}

Mergers appear to be commonly occurring in galaxy clusters based on the 
large amount substructure seen in their X-ray and optical morphologies. Signatures of active
merger are seen directly in temperature maps prepared from X-ray observations or indirectly in 
non-thermal radio and X-ray emission from cosmic-ray electrons accelerated at the shock front
formed by the merger.
A merger in Abell 754 was inferred from modeling the
integrated HEAO-1 \& 2 X-ray spectrum with a two temperature components in the approximate
temperature ration of 3:1 (Henriksen 1993).
 {\it ASCA} observations
resolved the spectral components into a temperature map with modest resolution (5 armin square regions
in the central region of the cluster). Some
regions that were hotter than the average cluster temperature and
inferred to be shocked gas from the merger. Comparison of the temperature
map to
simulations suggested 
of a major, off-center merger (Henriksen \& Markevich 1996). 
Similar temperature structure was also seen in the {\it ROSAT} temperature map (Henry \& Briel 1995).
Optical observations showed subclustering in the galaxy distribution
providing further evidence for the merger (Zabludoff \&
Zuritzky 1995). The possibility
that Abell 754 has had several mergers has been raised by Markevitch et al. (2003). 
Low frequency radio observations (Kassim et al. 2000) have uncovered two regions of extended radio
emission with an estimated spectral index that is consistent with a steep spectrum source. The
radio observations show that the western radio halo is spatially located at the hottest cluster region
seen by {\it ASCA} suggesting a connection between the shocked gas and cosmic-ray acceleration. 
Simulations of cosmic-ray acceleration in a cosmological setting (Miniati et al. 2001) 
predict that the non-thermal X-ray and radio emission should be contemporaneous.
The non-thermal X-ray emission
is difficult to detect because the powerlaw is steep and therefore provides considerably less emission
in the hard X-ray than in the soft X-ray. 
In the hard X-ray ($>$25 keV), though the spectrum should be free of the thermal cluster
emission, only large field-of-view, non-imaging detectors. Thus it is
necessary to account for the effect of 
AGN contaminating the cluster non-thermal emission. In the soft X-ray, one must accurately model the thermal components
which are generally not single phase in the cluster center, in order to measure the non-thermal component.

Cluster spectra are often better fit by two components.
This result is found for a number of clusters and across a wide range of data sets, including those
with low spatial resolution: IC 1262, Abell 3560, Abell 85, {\it BeppoSax} (Hudson, Henriksen,
\& Colafrancesco 2003; Bardelli et al. 2002; Neto, Pislar, and Bagchi 2001); three poor clusters,
{\it ROSAT} (Dahlem and Thiering 2000); Coma, Abell 2319, and Abell 2256, {\it RXTE} \& {\it ASCA} 
(Gruber \& Rephaeli 2002; Rephaeli \& Gruber 2002; Henriksen 1999); Abell 1367, {\it RXTE} \& {\it ROSAT} (Henriksen \&
Mushotzky 2001); the Hydra A and Centaurus clusters, {\it ROSAT} \& {\it ASCA} (Ikebe et al. 1997; Ikebe et al.1999).
Two component spectra are found in clusters with
a wide range of characteristics 
including: clusters with both regular and irregular optical morphologies,
clusters that appear to be undergoing major mergers or minor
accretions, and those both with and without radio halos or relics.
Two component spectral models consist of two thermal or a single thermal
plus a non-thermal component. For some clusters, two
thermal components are clearly preferred statistically over a non-thermal model,
while for other clusters one must rely
on physical arguments to determine the correct model.

In this paper, we model the {\it ROSAT}, {\it BeppoSax}, and Chandra
X-ray observations of the Abell
754 galaxy cluster with thermal and non-thermal models that are well 
matched to their spatial and spectral capabilities. We will
present evidence for a new, low temperature component in this
cluster and compare its measurements to our simulations of the
emission from embedded galaxy groups.


\section{Observations and Analysis}

Abell 754 is a rich Galaxy Cluster located at (J2000) 09$^{h}$08$^{m}$50.1$^{s}$,-09$^{\circ}$38$^{m}$12$^{s}$ 
with redshift of 0.0542 (Struble \& Rood 1999).  It was observed by Chandra on October 30, 1999 for 44,700 seconds.  
The peak cluster emission is located on chip 1 of the ACIS-I detector and the elongated emission spans all four chips.  
The total count rate of the uncleaned data is 18.35 $\pm$ 0.02 cts s$^{-1}$ with peak of 51 $\pm$ 2 cts s$^{-1}$, which
 is well below the saturation limit of the ACIS chips operating in Faint Mode (170 cts s$^{-1}$).  Figure 1 is a color
  coded intensity map that shows the full 21$\arcmin \times$ 21$\arcmin$ image in the 0.3-8.0 keV band.  The image was
   obtained using processing routines outlined in the CIAO 2.2.1 Science Threads.  Since A754 was observed during a
    period in which the ACIS temperature was -110$^{\circ}$C, it was important to correct the data for the increased 
    Charge Transfer Inefficiency (CTI). This was was done using the scripts provided at the CXC web-site.  The data 
    was then cleaned for flares and point sources.  A background map was created following the CIAO 2.2.1 Science 
    Thread {\it ACIS Blank-Sky Files} which uses as its basis a blank-sky datasets.
    The {\it merge\_all} script was used to create a exposure map and exposure corrected source image.  The 
    background was normalized to the exposure time of the data set, divided by the exposure map and then subtracted 
    from the exposure corrected source image.  The background subtracted, exposure corrected image was then smoothed 
    to form the image seen in Figure 1. 

The intensity map is quite complex overall and shows several interesting features. The central peak is 
elongated along the NE and SW directions.  The diffuse emission drops off sharply to the East and NE of 
the elongated central peak, but extends to the West past the edge of the ACIS-I Detector.  A small, 
second X-ray peak in the diffuse emission is visible on the western edge of the I3 chip.


Spectra were extracted from the cleaned, CTI corrected ACIS event files.  The redistribution matrix 
(RMF) for each region was determined by taking a count weighted average of the relevent RMFs provided 
by the CTI team.  
Since the RMF energy bins varied in size, the CIAO tool {\it mkwarf}, was unable to 
parse the RMF energies.  Therefore, the Ancillary Response Function (ARF) for a region was created by 
combining ARFs created using {\it mkwarf} for each energy range of identical energy bins.

In addition to the Chandra observation of A754, there are three 1996, near on-axis, {\it ROSAT} Position 
Sensitive Proportional Counter-B (PSPC) observations (total integrated time $\sim$16,500 seconds), 
and a May 17, 2000 {\it BeppoSax} observation with the Medium Energy Concentrator Spectrometer (MECS) ($\sim$121,000 seconds) 
and the Phoswich Detection System (PDS) ($\sim$55,000 seconds).  The PSPC and MECS data were extracted from regions 
centered on the {\it BeppoSax} pointing (09$^{h}$09$^{m}$21.6$^{s}$,-09$^{\circ}$41$^{m}$02$^{s}$) to facillitate
creating proper response matrices and producing proper background PHAs.

The PSPC data was reduced using standard FTOOLS 5.2.  First the gain was corrected using the {\it pcecor} 
inflight correction algorithm.  
The background was extracted from regions far from the source emission.  Care was taken to avoid regions 
obstructed by support 
structures.  The exposure map was used to correct for the differences in response of the PSPC for the 
source region and the 
background regions.  ARFs were created for each PSPC observation using {\it pcarf}.  The three source
 and background PHA files for each region were separately summed using {\it mathpha}.  The ARFs for 
 the merged spectra were created by using {\it addarf} to calculate a count weighted average of the 
 three individual ARFs from each period of observation.  The RMF for the gain corrected data was obtained 
from the HEASARC data base.  Overall, 15 PHA files were extracted to produce 5 merged PHA files: a merged 
PHA for an 8$\arcmin$ 
circular region centered on the {\it BeppoSax} pointing and four merged PHAs for four concentric regions spanning 
10$\arcmin$ in 2.5$\arcmin$ intervals.  The PDS's lack of spatial information and the inability to indentify 
point sources in the MECS (due to its relatively large Point Spread Function (PSF)) precludes the removal 
of point sources, therefore, point sources were not removed from the PSPC data in order to preserve consistency 
between the PSPC and {\it BeppoSax} spectral files. 

The MECS has a broad energy band and is used to constrain the diffuse thermal component.
An on-axis, circular region of radius with an 8$\arcmin$ radius  was extracted
from the merged MECS23 event file provided from the Italian Space Agency (ASI) archive.  ASI also provides 
an on-axis RMF and an 8$\arcmin$ background PHA file (taken from blank-sky fields at high galactic latitude) 
for the MECS23.  In order to create an ARF for our 8$\arcmin$ region of extended emission we 
 approximated the X-ray emission as radially symmetric. The SAXDAS program, {\it effarea}, uses a point source
  ARF and a surface brightness profile to produce a MECS ARF for a concentric, on-axis region of radially
   symmetric, extended emission (Fiore, Guainazzi, \& Grandi 1999). 
    An on-axis, point source ARF was obtained from the ASI web-site and the {\it ROSAT} High Resolution Imager 
    (HRI) data was used to create the surface brightness profile.  Since {\it effarea} only creates 
    an ARF for a single detector, 
    the MECS23 ARF was created by combining the ARFs created for the MECS2 and MECS3 detector.

Extracting concentric regions for the MECS data is difficult because extraction must be done in SKY 
coordinates in order to match the MECS regions with the PSPC regions.  However, since the background 
events file is from multiple observations, it is not possible to extract a background PHA file in SKY 
coordinates from the background events files provided at the ASI web-site.  We, therefore, constructed 
an algorithm that uses the MECS2 source data to convert DET coordinates to SKY coordinates.  This algorithm 
produced  an error of no more 3 pixel sky coordinates for the MECS2 source data, corresponding to an error 
of $\leq$0.4$\arcmin$, which is much smaller than the MECS PR$_{50}$ of 1.2$\arcmin$ at 6.4 keV. 
 Unfortunately we were unable to construct a similar alogrithm for the MECS3 data.  Four concentric 
 regions identical to the PSPC regions were extracted from both the source MECS2 events file and corrected 
 MECS2 background events file.  {\it effarea} was then used to create ARFs for each region, with the procedure described above.

PDS events files for A754 are currently unavailable, so the cleaned, background subtracted PHA file was 
obtained from the ASI archive.  
We also obtained the proper PDS RMF from the ASI web-site. Since the PDS is not an imaging instrument, it 
impossible to limit it field 
of view or remove point sources from it.  Significant point source contamination to the PDS spectrum of A754 comes 
from is 26W20.  
This  X-ray bright, 
radio Galaxy is visible in the PSPC at a distance of 24.5$\arcmin$ to the SW of A754's center, well within the PDS's field
 of view. 
Silverman, Harris, \& Junor (1998) report a total radio flux of 435$\pm$40 mJy emitted by core, tail, and lobe components. 
These authors
report an X-ray luminosity measured by the on-axis {\it ROSAT} HRI observation of 3.70 $\pm$0.10 $\times$10$^{43}$ ergs s$^{-1}$.The
 best fit model
for the emission using the {\it ROSAT} PSPC is $\alpha$ = 1.32 $\pm$ 0.17. 
Since 26W20 is well outside our PSPC and MECS extraction regions, it does not appear in their spectra, therefore a 
power-law component, 
with a photon index forzen at $\alpha_{x}$ = 1.3 and normalization 2.0$\times$10$^{-4}$ was included in all of the PDS spectral models, 
but excluded from the 
spectral models of the other detectors.  

\section{Chandra Temperature and Surface Brightness Map}

A map of the optical galaxy structure (Figure 1) was prepared for comparison with X-ray and radio structure.
The galaxy sky surface density was determined for the region surrounding A754.  The Digitized Sky Survey (DSS) 
image of a single Second Epoch Survey R-band plate of the region\footnote{{\tt http://archive.stsci.edu/cgi-bin/dss\_plate\_finder}}
were processed by the source extraction tool, 
{\sc SExtractor}
 The tool produces a listing of object positions, fluxes, elongations, and FWHMs among other parameters.  
 The parameters are used to discriminate between galaxies and stars, with galaxies being associated with any object that is 
 particularly dispersed or elongated. The galaxies were selected down to a limiting magnitude of
 19. The threshold for being considered dispersed was a FWHM $>$ $4\arcsec$ 
 (the FWHM of stellar images is $3.0\arcsec$) while an elongation threshold of the long axis being 1.4 times larger than the 
 short was used. Note that the PSF for the plate is slightly elongated, presumably due to tracking errors.  
 The positions of the extracted galaxies were smoothed by an adaptive kernel to produce a number-density field.  
 For each point in the field, the adaptive kernel maintained a constant number of 16 galaxies in the density estimate.  
 The field was then smoothed on a scale, $41 \arcsec$, equivalent to 50 kpc at the distance of the cluster.
Iso-contours are at levels 2, 4, 6, ... $\sigma$ above the background levels.

The temperature map (see Figure 2, region values in Table 1) shows two important features that were previously identified in the 
temperature map obtained with {\it ASCA} (Henriksen \& Markevitch, 1996): hot, ~16 - 20 keV gas (region 2)
near the NW galaxy component and cooler, $\sim$7 keV gas North
of the SE galaxy component and co-spatial with the X-ray peak. 
There is a gradient in {\it ASCA} temperature map running E-W approximately
through {\it Chandra} regions 5-14-1-3-2 that is also visible in the {\it Chandra} map.
A radial temperature profile was made from the {\it Chandra} 
image with the center on the X-ray peak (see Figure 3). The radial profile becomes dominated
by gas to the west of center at radii greater than $\sim$4$\arcmin$, and clearly confirms that
the temperature increases across the cluster from east to west by nearly 50\%
similar to the ridge structure visible in the {\it Chandra}
and {\it ASCA} temperature maps. 

The cooler regions are consistent with the temperatures for those regions 
measured with {\it ASCA}. However, the {\it Chandra} map shows that the cooler gas is 
spatially coincident with a sharply delineated "peanut shaped region"
in the surface brightness map. The low frequency radio image (Kassim et al. 2001), shows two
distinct, high intensity regions: the small radio region to the east of the X-ray 
peanut, and the large radio region that lies to the West. 
The peanut appears to be a cool, low pressure region sandwiched between the radio emitting
regions. We propose that the peanut morphology is formed by the pressure exerted on it
as the surrounding radio regions expand into it. The radio regions are hotter
than the surrounding cluster atmosphere which is consistent with the
hypothesis that cosmic-ray acceleration occurs at the
shock front. Under the simplest assumption that primary cosmic-ray electrons are
accelerated at the shock front and then propogate out until they radiate 
their kinetic energy via the inverse-Compton process, the electrons do not travel very far, $\sim$10 kpc.
Thus, the proximity of the radio emission to the shocked gas is consistent with
the simplest model of non-thermal emission by primary cosmic-rays. 
Miniati et al. (2001) found that the proton cosmic-ray pressure
may be up to 45\% of the thermal gas pressure in clusters. Thus, the proton pressure
may be an important contribution to the dynamics of the cluster atmosphere when expanding into gas that is significantly
cooler than the ambient cluster gas.

\section{Abundance Map and Radial Profile}

The abundance map derived from fitting the MECS and PSPC radial profiles is shown in Figure 4.
Combining the MECS and PSPC provides better azimuthal coverage of the cluster and better
contrains the radial abundance than does the {\it Chandra} observation.
The radial profile is generally consistent with a constant abundance value
of 30\% Solar, the global value reported in Table 2. 
The 90\% uncertainties of outer two abundance measurements points do
not quite overlap suggesting that higher abundance gas may have been dumped during the merger. There is
no gradient near the X-ray peak. 
Monatonically decreasing metallicity gradients typically anti-correlate
with cluster mergers (De Grandi \& Molendi 2001). Thus, a merger acts to mix the gas and erase
this type of pre-existing gradient. In the case of Abell 3266 (Henriksen \& Tittley 2002), an off-center abundance enhancement
was found, as opposed to a gradient. In Abell 3266, it appears from the geometry of the merger, that the
enhancement results from deposition of higher abundance gas during the merger. Similarly, the outer annular region
that shows an abundance enhancement in Abell 754 could also be due to deposition of higher density
gas during the merger. An off-center abundance profile is more difficult to detect with high significance
because the emission integral $\sim$n$^{2}$ and the gas density typically is $\sim$n$^{-2}$ outside of the
cluster core. Thus the emission integral at 1 Mpc is a factor of
$\sim$250 lower than in the center (for an equal volume). Nevertheless, off-center abundance profiles as
in the Abell 3266 and Abell 754 may be
important diagnostics for cluster evolution, providing clues to the clusters merger history.
Firthermore, formation of a metallicity gradient in galaxy clusters may depend more
on the presence of a central, cD galaxy than on cluster richness (Umetsu \& Hattori 2000). The
cD in Abell 754 lies near the NW galaxy clump whereas the X-ray peak is near the SE galaxy component. 
his large spatial offset
and proximity of the cD galaxy to the NW galaxy peak suggests that the cD is identified with the subcluster, located
near the hottest region, and may be identified with the high abundance at large readius.

\section{Soft X-ray Emission From Embedded Galaxy Groups}

A large region of the {\it BeppoSax} MECS and {\it ROSAT} PSPC image was selected
for analysis to search for non-thermal X-ray emission. These instruments provide broad band coverage that is
able to constrain the cluster components: 0.5 - 2.0 keV 
(PSPC), 1.5 - 10.0 (MECS), and 15 - 200 keV (PDS). Together they offer
a broad energy band of 0.5 - 200 keV with reasonable signal-to-noise. The MECS has the
best hard X-ray coverage of the recent imaging instruments. For comparison, while the {\it ASCA} GIS
has a comparable bandwidth, the MECS has a factor of 2 higher effective area at 8 keV. For a 
hot cluster such as Abell 754, the thermal components will dominate these spectra and the
PDS is crucial for detecting the non-thermal component at high energy. However, the PDS
is most sensitive to flat spectrum sources which tend to be AGN rather
than diffuse radio emision. The PSPC may be sensitive 
to the steeper powerlaw that characterizes diffuse non-thermal emission; a component
that may dominate the spectrum at low energy. 

The results obtained from fitting these data with various single and
multiple thermal and non-thermal models are given in Table 2. The sequence
of Figures 5 - 8 show data, model, and residuals for progressively more complex
models. Figure 5 shows a single mekal model fit to
the data. Significant residual emission appears around 100 keV and below 1 keV. The PDS
field of view contains the BL LAC object, 26W20.

Modeling 26W20 in the PDS with the parameters from Silverman, Harris, \& Junor, (1998) removes the high energy residuals
(Figure 6). The remaining soft residuals can be modeled by adding either a
non-thermal component (Figure 7) or
a second thermal component (Figure 8). The parameters determined from these fits are given in Table 2.
The second, diffuse emission component may either be thermal,
with temperature 0.75 - 1.03 keV and
7.0$\times$10$^{43}$) ergs s$^{-1}$ luminosity which is 7.2\% of the the hot thermal
component, or non-thermal with 
$\alpha \simeq$2.3 (See Figure 9) and bolometric (0.1 - 100 keV) X-ray luminosity of 2.4$\times$10$^{42}$ ergs s$^{-1}$ or 0.25\% of 
the hot thermal component.  The detection of non-thermal emission is 3.65$\sigma$ 
with a significance of 99.7\% using the ftest. However, as can be seen in
in Table 3, while the additional components both reduce the total number of degrees of freedom
by 2, the thermal reduces $\chi^{2}$ by 53 compared to 13.5 for the non-thermal. So, the thermal component
provides the best fit to the data. As noted earlier, the X-ray contour map shows a 
possible source in the hottest region of the temperature map, region 2. We fit this region
with a powerlaw to compare to the thermal model to test if this region is the source
of a non-thermal component. The thermal component is statistically a much better fit to
the data. A powerlaw only gives a comparable fit to the thermal model if the column
density is allowed to be three times the Galactic value. In this case, the spectral index
is 1.5-1.55, comparable to AGN. As there are not strong FeK$_{\alpha}$residuals apparent in the
powerlaw fit, we can not rule out that the region is dominated by
an absorbed powerlaw source rather than hot gas.

There are a number of possible sources for cool thermal gas including: galaxies, groups, and
intercluster gas. Proposed sources of a low temperature 
thermal component of galactic origin include the integrated
emission from gas associated with several elliptical galaxies or groups within the observed region of the cluster 
(Henriksen \& Silk 1994). A number of small scale structures exist in the Abell 1367 cluster that are significantly
cooler (0.3 - 0.9 keV than their surroundings (3 - 4.5 keV) (Sun \& Murray  2002) 
suggesting that the early-type galaxy coronae can survive in the intracluster
medium. There are also several similar compact sources not identified with galaxies in that cluster.
The central dominant ellitpical galaxies in the central region of the Coma cluster retain their
corona and have a temperature of $\sim$1 keV (Vikhlinin et al. (2001). We will perform simulations
that will test this hypothesis by making a simulated temperature map for comparison to our Chandra temperature map.
The {\it Chandra} temperature map maintains approximately
6000 counts per region to allow spectral fitting of each region. The smallest regions
are then 1 x 1 arcmin and the largest are 2 x 2 arcmin. At the redshift of Abell 754, 1 arcmin = 73 kpc (H$_{0}$ = 65 km s$^{-1}$ Mpc$^{-1}$).

Simulations were performed of the emission weighted 
temperature from embedded groups with pixels of similar size to the pixels in a Chandra temperature map.
 The simulations assume X-ray emission from identical groups (e.g., similar temperature (0.88 keV) and
luminosity (the total cool component/number of 
groups).
The groups are embedded randomly in a region 876 kpc x 876 kpc x 1000 kpc which approximates
the region of the {\it BeppoSax} analysis that yielded the cool component.
The beta value of the surface brightness profile is fixed at 0.5 and the core radius is allowed to span a range of values 
that includes groups with very peaked and very flat distributions. The central density for the groups
is determined from the luminosity within 250 kpc, a typical radius observed for group emission (Mulchaey et al. 1994). The luminosity for a group
is observationally constrained since it must be 
equal the total cool component luminosity divided by the number of groups.
The Abell 754 cluster gas parameters are n$_{c}$ = 
1.85$\times$10$^{-3}$ (Abramopoulos \& Ku 1981), core radius = 300 kpc, and kT = 10 keV. 
We simulated the temperature map with 2 or more groups embedded in the cluster's central region. The groups
range from small, ~10 kpc core radius to large, ~100 kpc. While groups are observed with larger core radii, the
effect of lowering the observed temperature becomes less (see Figure 10) and so there was no need to  investigate
larger core radii groups. We also did not look at many small halos, $>$20, because this becomes unphysical for the following reason.
Groups with significant X-ray emision contain at least one elliptical galaxy. Thus, many small groups mean many elliptical galaxies.
As discussed below, elliptical galaxies alone are not luminous enough and the sources
giving the cool component. If one requires that the ellipticals have a luminosity comparable to groups,
then the luminosity function is violated.
Embedding a single, luminous, cool group or several, less luminous groups will change the 
emission weigthed temperature of the regions that the groups are in. Large core radii groups
may affect more than one pixel and groups located further from the center of the cluster will
have a more significant impact on the emission weighted temperatures in the temperature map.
Figure 10 shows the lowest temperature of any pixel in the simulated temperature map for the range of group core radii
and number of groups tested.
We find that for the full range of core radii and number of groups, the lowest temperature observed for any pixel in the
simulated map is well above the observed minimum in the {\it Chandra} map, 6.9 keV. 
Thus, a model with either a small number of luminous groups or many low luminosity groups can reproduce the the cool component and 
not violate the Chandra temperature map.
The constraint from the temperature map on the simulations is that no pixel has a lower emission weighted temperature
than 6.9 keV, with 90\% confidence, within 600 kpc (the radius of the cool component region).
Thus we conclude that groups are a plausible source of the cool component. 

Elliptical galaxies have a hot gaseous halo at 1 keV, comparable to groups, but also typically 
contain less gas. A King model for Abell 754 parameters (Abramopoulos \& Ku 1983)
using the central surface density of galaxies of 92 Mpc$^{2}$ and the 
X-ray core radius for a King model (0.71 Mpc) estimates that there would be 52 galaxies within 
the 400 kpc region. A typical elliptical fraction is 80\%.
This would give 42 elliptical galaxies. To give the cool component would require and average L$_x$ of 1.7$\times$10$^{42}$ ergs s$^-1$ per
elliptical galaxy. The distribution of normal early-type galaxies (Eskridge, Fabbiano, \& Kim 1995) show that 10$^{41}$ ergs s$^{-1}$
is typical and that only a small fraction have X-ray luminosities a factor of 10 higher. Thus it is implausible that
this large cool component is from elliptical galaxies only. 

Another possibility is a diffuse
baryonic halo surrounding the cluster. 
Evidence for a similar, diffuse component in the 0.5 - 1 keV  range have been reported with several different data sets and authors.
Henriksen \& White (1966)
using HEAO-1 and Einstein data with a similar broad band coverage but with much lower effective area reported 0.5 - 1 keV
gas beyond the central cooling flow region in several clusters and showed that the amount of the cool component
did not correlate with the amount of cool central gas and is therefore not "spillover" from the cooling flow region
that is outside of the SSS field-of-view.
Kaastra et al. (2003) report a soft X-ray component that is visible as an excess in the 0.4 - 0.5 keV range that
is attributed to the warm hot IGM.
Bonamente, Joy, and Lieu (2003) report evidence for
a diffuse baryonic halo around the Coma cluster that extends to 2.8 Mpc. Their spectral analysis of {\it ROSAT} PSPC observations
show that the radial temprature profile of this component is consistently 0.25 keV with 10\% uncertainty. This suggests that
the intercluster component is substantially cooler than the Abell 754 component since out temperature
of our component is consistently higher than this. Nevalainen et al. (2003) report
soft excesses in XMM observations of the Coma, Abell 2199, Abell 1795, and Abell 3112 in the cluster's inner 0.5 Mpc, a similar 
region to that covered by our MECS/PSPC analysis. The characteristic
temperature is 0.6 - 1.3 keV, which is also consistent with the Abell 754 cool component. They point out that their
 density, 10$^{-4}$ - 10$^{-3}$
cm$^{-3}$ is well above those expected from the WHIGM (Dave 2001) projected into the core region of the cluster.
The density of the Abell 754 cool component is similarily high compared to the WHIGM. The PSPC observations
provide some spatial information information on the soft component distribution. Figure 11 shows a radial
surface brightness distribution of the cool, hot, and galaxy components. The emission is slighltly peaked in the center
relative to the hot component. The {\it BeppoSax} pointing is used
as the center of the distributions and lies slightly to the west of the X-ray peak.
The galaxy distribution from the DSS is shown for comparison. It is interesting that the soft X-ray emission is not
flat, as would be expected if it were projected emission from filamentary gas surrounding the cluster. Also, the
highest region of soft emission is located between the regions of radio emission. In the simplest model for
non-thermal emission, synchrotron-inverse Compton from primary cosmic rays, the non-thermal radio and X-ray should
be co-spatial. Thus, if the soft component is non-thermal, that would require
that the magnetic field is very non-homogeneous in the cluster so that the  
visible radio regions are simply where the magnetic field is highest.
This interpretation is somewhat unsatifactory because, if the magnetic field os
"frozen in" to the gas, then the region with compressed gas, in the center, must also have the highest magnetic field. 
This is the opposite of what is required by the location of the radio sourse.
In the absence of
the non-thermal X-ray emission, one could explain the absence of radio emission in the center as synchrotron aging 
due to a higher magnetic field in the high density, central region.
But non-thermal X-ray requires that relativistic electrons are present and simulations show that radio emission 
should be contemporaneous with the non-thermal X-ray emission.
Based on our simulations, the best explanantion for the soft component is 0.5 - 1 keV 
emission from galaxy groups embedded within the hot ICM.

\section{Cluster Merger History and Non-thermal Radio Emission}


The results of our analysis of the various X-ray data sets brings together new details of a dramatic cluster merger
in Abell 754.Earlier, we presented evidence of temperature structure in the intracluster medium that we attribute
to shock heating during a merger. The radio structure of Abell 754 is complex in that 
there are two extended features near the X-ray and optical
cluster centers. The western radio halo appears to be the result of shock acceleration at the shock front during a major merger as it is
coincident with the hottest gas regions in the cluster. 

Simulations of cluster merger (Takizawa \& Naito 2000) trace the spatial distribution and time
evolution of synchrotron radiation
due to electrons accelerated at the shock fronts that form during a cluster merger. The morphology of the radio halo
and its observed location relative to the shocked cluster gas depends on the viewing angle relative to
the merger axis and the age of the merger. When viewing perpendicular to the merger axis, the radio and
thermal X-ray have different spatial relationships at different times in the merger. The shocked gas that is
west of the cluster center is co-spatial with the radio halo. This morphology seems to be consistent with the morphology
after maximum contraction. The simulations show two outward traveling shocks with the radio and shocked gas co-spatial. 
While viewing nearly along the line of sight could
also account for the shocked gas and radio being co-spatial, that viewing angle would also make both regions appear
to be in the center of the cluster, which is not the case since they are west of center. Thus we conclude that
the merger is in the plane of the sky or at fairly large angle to the line of sight.

The eastern radio source is not coincident with shocked gas and requires a more complex evolutionary scenario. Since the 
X-ray peak near the eastern radio source does not appear to be shocked, it is 
consistent with the simulations of an off-center merger (Roettiger, Stone, \&
Mushotzky, 1998) that was hypothesized based on the {\it ASCA} temperature map (Henriksen \& Markevitch 1996). A band of cool
gas is produced running through the center in this scenario. This gas is a mixture of pre-shocked primary cluster gas and cooler subcluster
gas in the simulations and is seen in the temperature map. The Eastern radio source may then be from an earlier merger or accretion event
for the following reason. 
Since radio halos loose their energy due to inverse-Compton cooling 
and have a relatively short lifetime compared
to the non-thermal hard X-ray component 
(Takizawa \& Naito 2000), the radio halo fades quickly after the most contracting epoch, 
when the magnetic field is strongest, during
the cluster merger. 
Thus, it would be unlikely for the radio source to still be around after the
thermal signs of the merger are gone in the simplest model. However while it is unlikely, Ohno, Takizawa, \& Shibata (2002)
find that turbulent Alfven waves reaccelerate cosmic-rays that originated at the shock front thereby extending
the lifetime of the radio halo. The turbulent gas mixing would also diminish the regions of hot gas from the shock front
gas. Alternatively, secondary electrons produced by proton-proton collisions followed by pion
decay may produce a radio halo with a longer lifetime (Blasi \& Colafrancesco 1999). Our hypothesis that
the western radio halo is due to primary electrons and the eastern source due to secondary electrons
can be tested by measuring the radio spectral index for each component.
The spectral index is predicted to be flatter for secondary electrons relative to primary
so that a comparsion of the spectral indeces is
a check on our hypothesis. 


For the non-thermal component, our best fit normalization and gamma, 2.3, gives 3.2$\times$10$^{14}$ ergs 
cm$^{-2}$ s$^{-1}$ keV$^{-1}$ @  20keV
which is lower than 90\% confidence upper limit reported by Valinia et al. (1999) using {\it RXTE},
7.1$\times$10$^{14}$ ergs cm$^{-2}$ s$^{-1}$ keV$^{-1}$ @  20keV. The upper value, 8.49$\times$10$^{14}$ ergs 
cm$^{-2}$ s$^{-1}$ keV$^{-1}$ @  20keV (90\% confidence),
slightly exceeds the upperlimit from the {\it RXTE}. However, under the simplest assumptions
 (e.g., homogenous magnetic field), the non-thermal X-ray
component and the radio halo emission should be co-spatial and the fact that the peak of the soft component
resides between the radio halos, as discussed in the previous section, 
makes a non-thermal origin of the soft X-ray component unlikely.

\section{Conclusions}

We report on the detection of a soft X-ray component in Abell 754 out to 600 kpc from the cluster center. The cool component
peaks near the X-ray center of the Abell 754 cluster and decreases with radius. A thermal model
is statistically a much better fit than a non-thermal one for the cool emission. In addition, the peak of the
soft component lies between the radio halos thus making a non-thermal interpretation
for the X-ray emission unlikely.

Several potential thermal sources have been evaluated: (1) galaxy coronae
embedded in the intracluster medium, (2) galaxy groups embedded in the intracluster medium,
and intercluster baryons. While the density and temperature may be too high to
match simulations of intercluster baryons, most importantly, the soft component
is peaked in the cluster center and therefore must be trapped in the cluster
potential well, thus excluding an intercluster origin. Typical galaxy coronae would be required to be an order of
magnitude more luminous than
is observed in surveys of early-type galaxies to give the low energy emission,
excluding them as a source of the emission. Groups match the temperature and total luminosity of the
low energy 
component. Our simulations show that either a single luminous group or several
lower luminosity groups can reproduce the low energy
component without violating the {\it Chandra} temperature
map constraints. This remains the most plausible explanation for the low temperature
emission.

\figcaption[old_DSS_R_ACIS_Overlay_colour.eps]
{Color coded intensity map of background subtracted image of the A754 cluster on the ACIS-I array.
Galaxy iso-contours are at levels 2, 4, 6, ... 
$\sigma$ above the background level.\label{fig1}}


\figcaption[A754_Tempmap.eps]{The 2-dimensional temperature map obtained from the
{\it Chandra} observation. Hot gas is visible to the West of the center
and cool gas running across the X-ray peak on the East side.\label{fig2}}

\figcaption[A754_30k_profile.eps]{Radial temperature profile is consistent
with the temperature map in showing that the temperature generally increases
across the cluster from East to West. There are 30,000 counts per region.\label{fig3}}


\figcaption[Abundance_profle.eps]{An abundance enhancement
is seen 8 arc min from the center of the main (SE) cluster perhaps
due to the subcluster involved in the merger.\label{fig4}}



\figcaption[mekal_plot.ps]{Thermal model shows residual hard and soft emission in the PDS. \label{fig5}}

\figcaption[mekal_w_src_plot.ps]{Thermal with powerlaw added to model contaminating source, 26W20 in the PDS. \label{fig6}}
\figcaption[mekal_pl_plot.ps]{Thermal and 2 powerlaws (a second added to model diffuse non-thermal emission). \label{fig7}}
\figcaption[mekal_mekal_plot.ps]{Two thermal models and point powerlaw. \label{fig8}}

\figcaption[spectral_index.ps]{Measured contours are consistent with
the spectral index measured for inverse-Compton emission produced by
primary cosmic-ray electrons in simulations.\label{fig9}}


\figcaption[MinT_5.eps]{Contour plot of minimum simulated temperature map region temperature
for combinations of group number versus core radius. The total emission from
the groups, in each case, is set equal to the soft X-ray component. Models with a few
luminous groups or many low luminosity groups are found to give simulated temperature
maps that do not violate the observed {\it Chandra} temperature map.\label{fig10}}

\figcaption[A754_SBP_norm2_logscale.eps]{The radial surface brightness of the cool (blue), hot (red),
and galaxy (black) components are shown. The cool component appears to be slightly
peaked in the central region relative to the hot component though and it is not flat
as would be expect for a projected intercluster component.\label{fig11}}

\begin{deluxetable}{lc}
\tabletypesize{\footnotesize}
\tablewidth{0pt}
\tablecolumns{2}
\tablecaption{Spectral Fits of Regions. \label{tbl-1}}
\tablehead{\colhead {Region} &\colhead{kT$_{1}$ (keV)}}
\startdata
 1 &12.7$^{+1.41}_{-1.16}$ \\ 
 2 &17.4$^{+1.85}_{-1.42}$ \\ 
 3 &16.7$^{+2.95}_{-2.3}$ \\ 
 4 &8.99$^{+0.85}_{-0.71}$ \\ 
 5 &9.6$^{+1.06}_{-0.86}$ \\ 
 6 &10.8$^{+1.28}_{-1.04}$ \\ 
 7 &11.2$^{+1.6}_{-1.24}$ \\ 
 8 &8.54$^{+1.11}_{-0.9}$ \\ 
 9 &11.6$^{+1.05}_{-0.82}$ \\ 
 10 &8.36$^{+1.74}_{-1.18}$ \\ 
 11 &16.7$^{+1.93}_{-1.46}$ \\ 
 12 &9.73$^{+1.11}_{-0.93}$ \\ 
 13 &6.76$^{+1.18}_{-0.91}$ \\ 
 14 &11.6$^{+1.44}_{-1.15}$ \\ 
 15 &7.49$^{+0.55}_{-0.46}$ \\ 
 16 &7.52$^{+0.69}_{-0.61}$ \\ 
 17 &8.66$^{+0.69}_{-0.59}$ \\ 
 18 &11.3$^{+1.28}_{-0.96}$ \\ 
 19 &4.56$^{+0.84}_{-0.65}$ \\ 
 20 &8.12$^{+1.06}_{-0.84}$ \\ 
 21 &8.7$^{+1.16}_{-0.94}$ \\ 
 22 &8.55$^{+0.61}_{-0.53}$ \\ 
\enddata
\end{deluxetable}
\clearpage


\clearpage

\begin{deluxetable}{lccccc}
\tabletypesize{\scriptsize}
\tablewidth{0pt}
\tablecolumns{6}
\tablecaption{Spectral Fits of 8$\arcmin$ Region. \label{tbl-2}}
\tablehead{\colhead{Model} &\colhead{kT$_{1}$ (keV)} & \colhead{Abundance} & \colhead
{kT$_{2}$ (keV)} & \colhead{$\Gamma_{X}$} & \colhead{$\chi^{2}$/dof}}
\startdata
MEKAL  & 10.08$^{+0.35}_{-0.33}$ & 0.31$^{+0.04}_{-0.04}$ &  & & 455.3120/383 \\
MEKAL + POWERLAW & 10.57$^{+0.93}_{-0.93}$ & 0.32$^{+0.10}_{-0.08}$ & & 2.28$^{+1.93}_{-1.31}$ &  441.8215/381\\
MEKAL + MEKAL & 10.53$^{+0.43}_{-0.40}$ & 0.31$^{+0.04}_{-0.04}$ & 0.88$^{+0.15}_{-0.13}$ & & 402.4572/381 \\
\enddata
\end{deluxetable}

\clearpage

\plotone{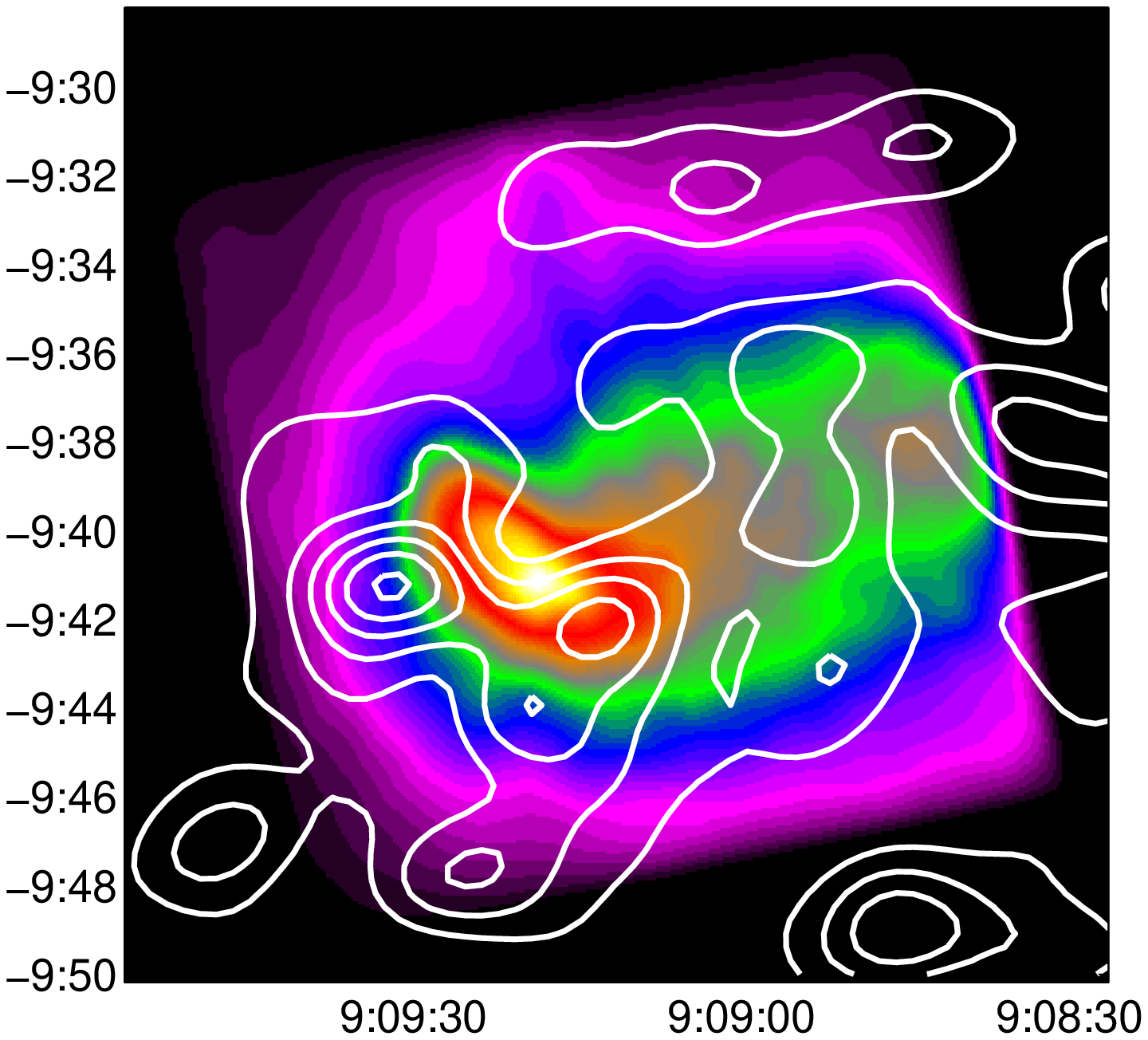}



\plotone{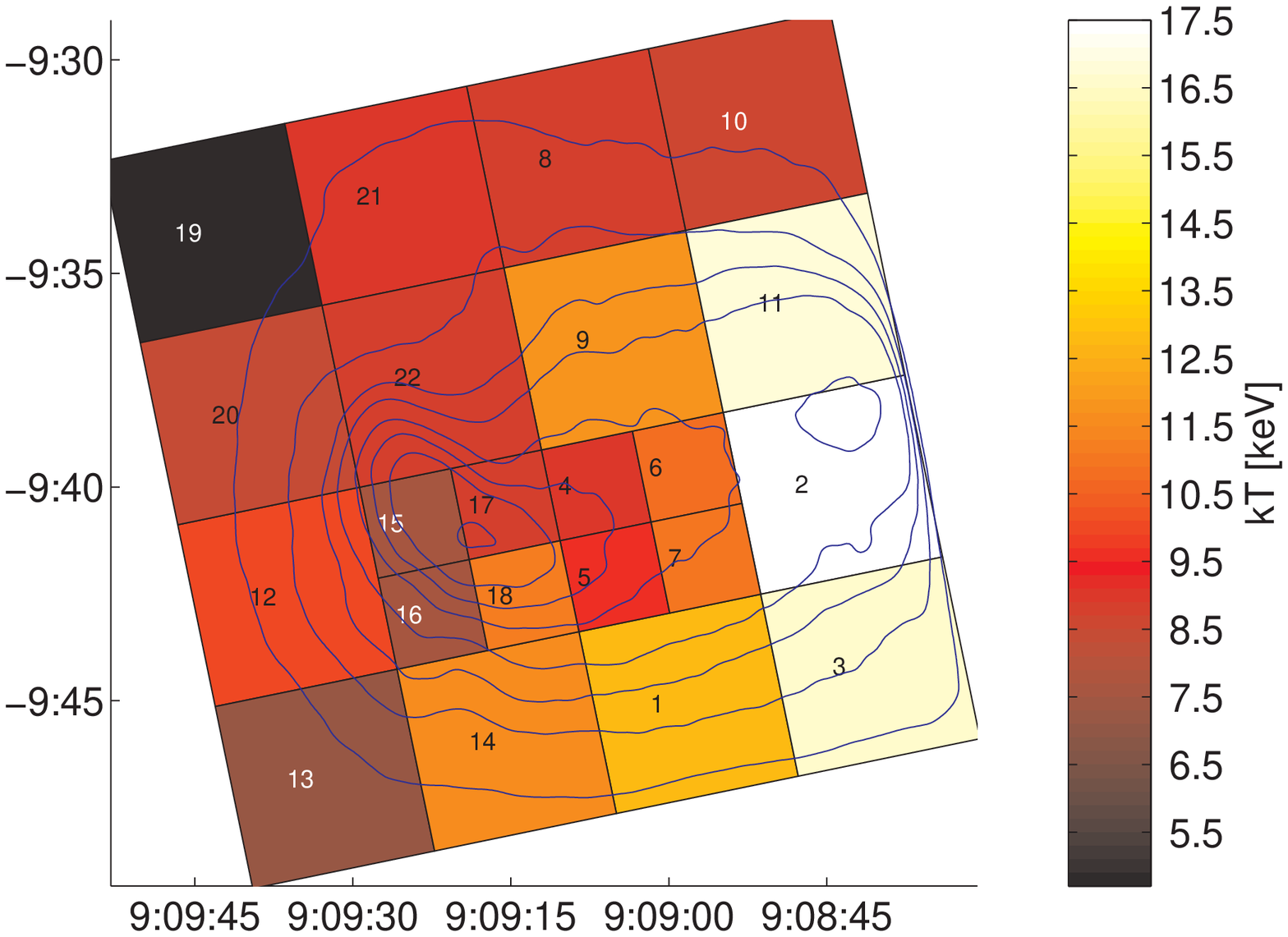}

\plotone{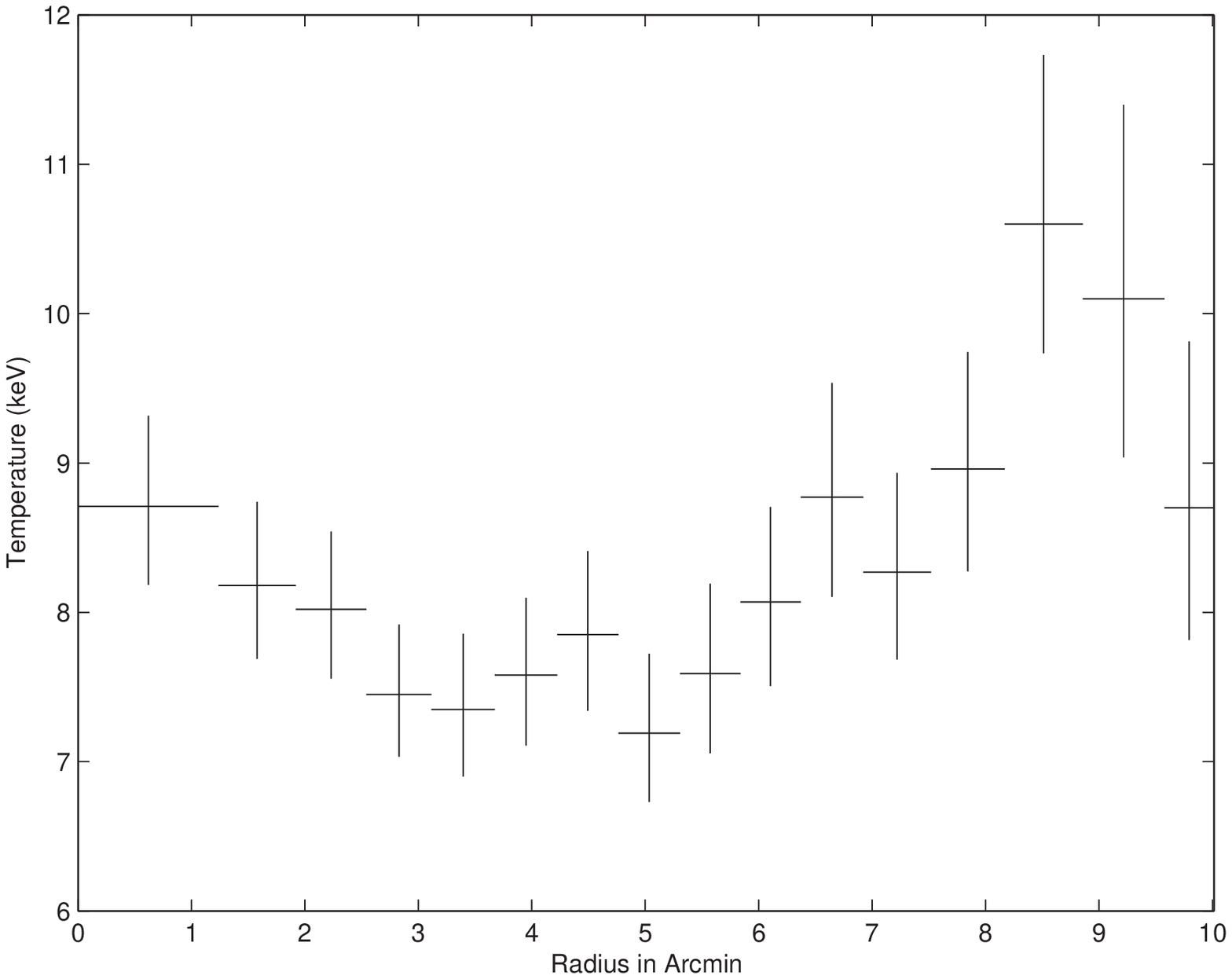}

\plotone{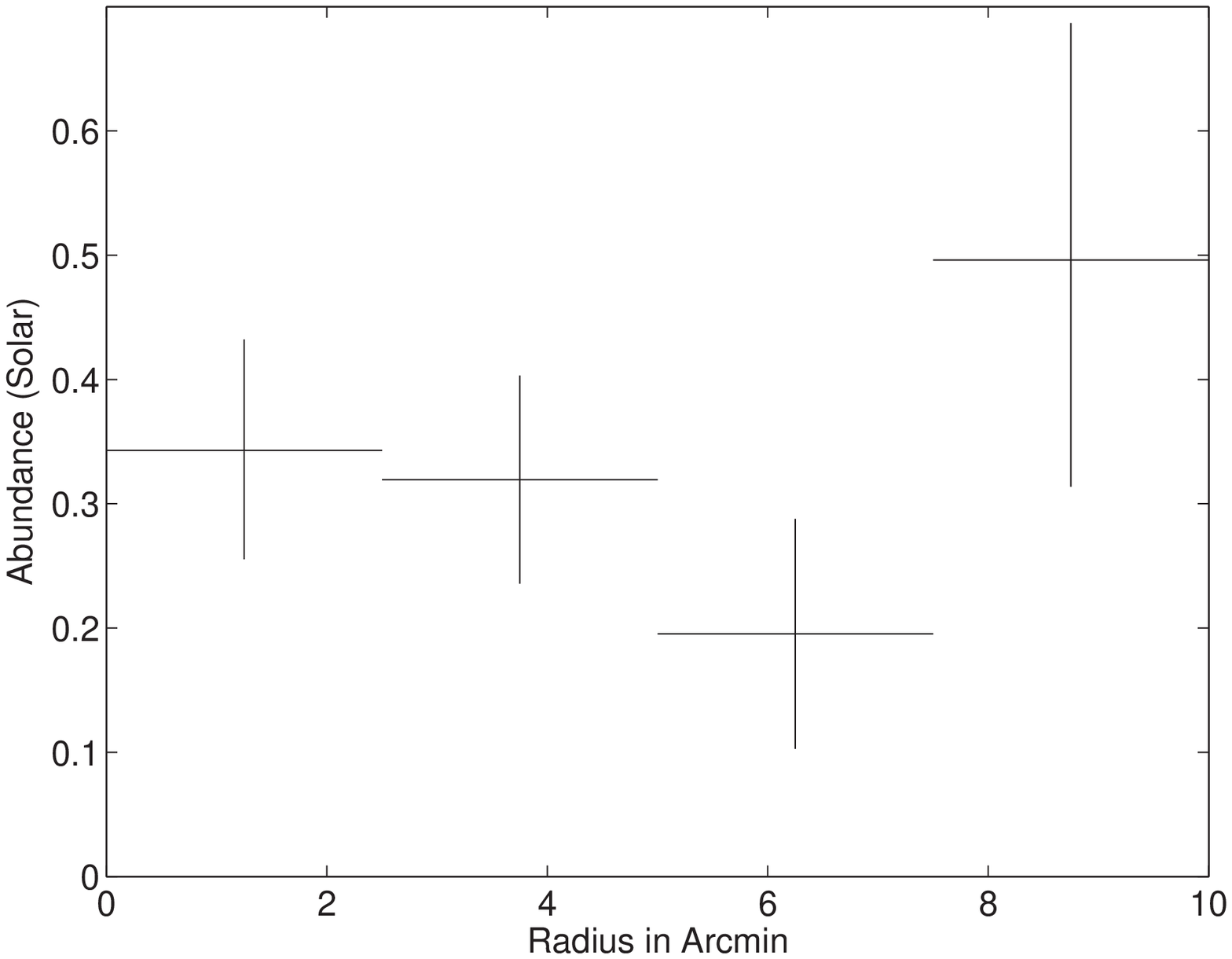}

\plotone{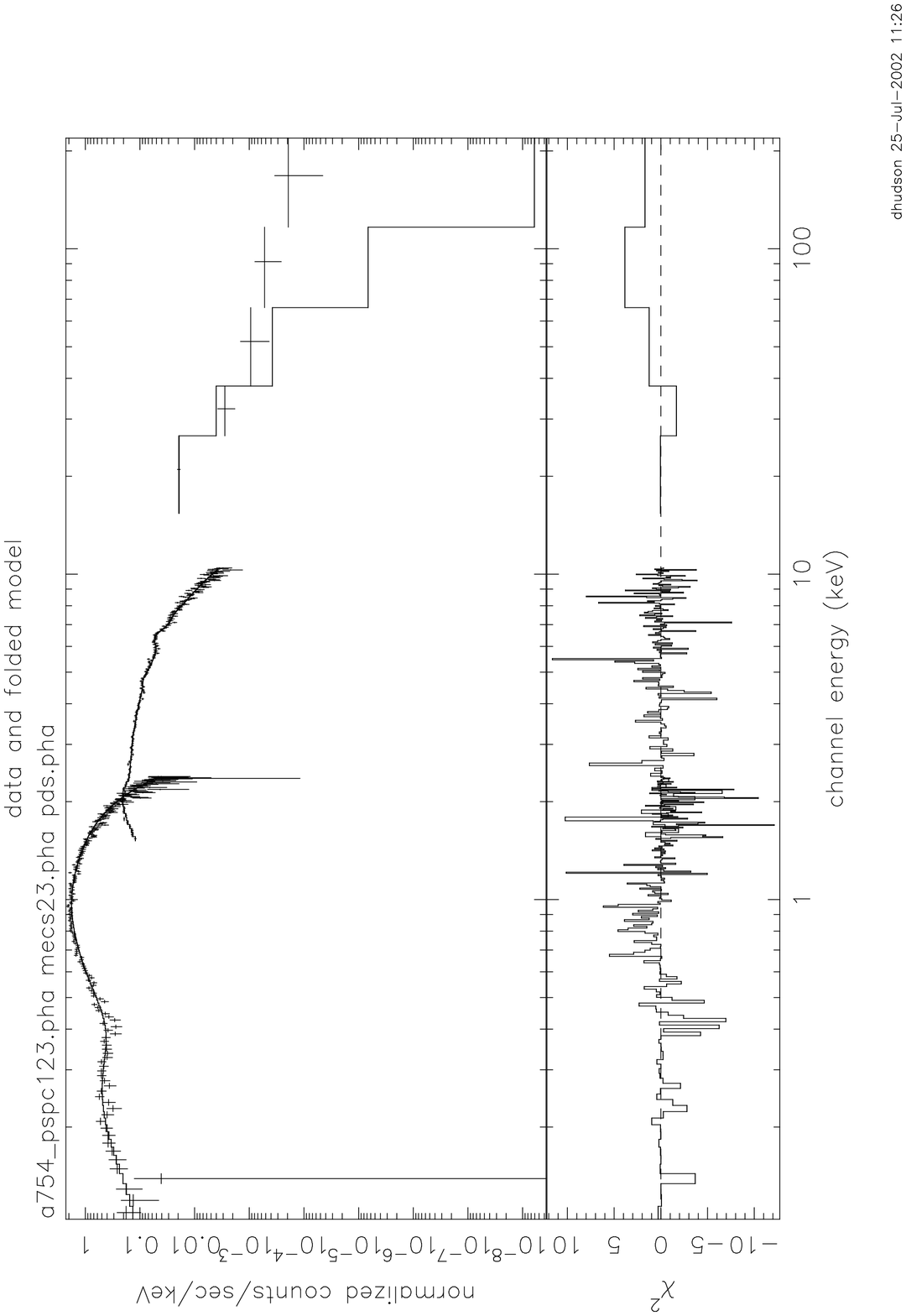}
\plotone{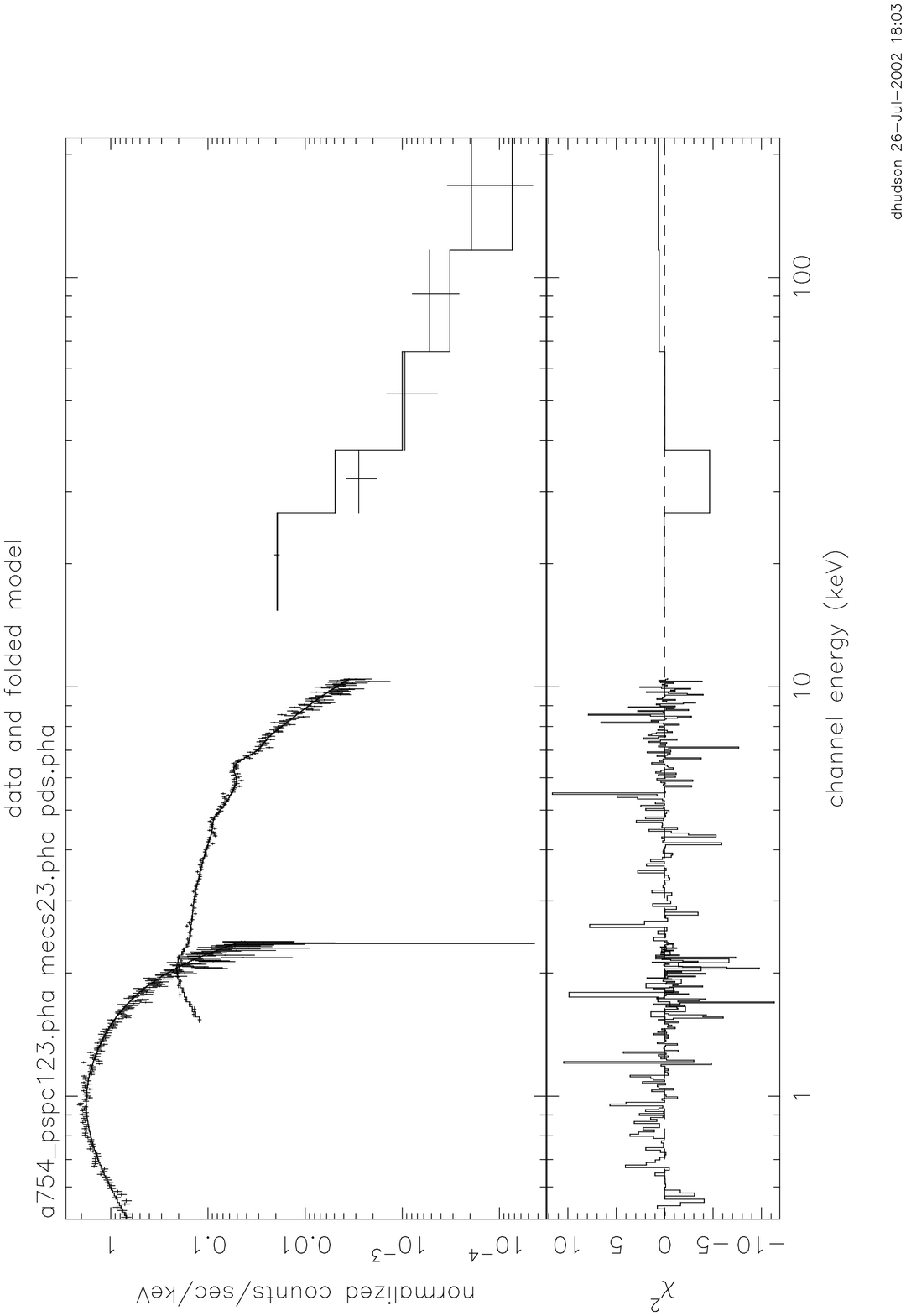}

\plotone{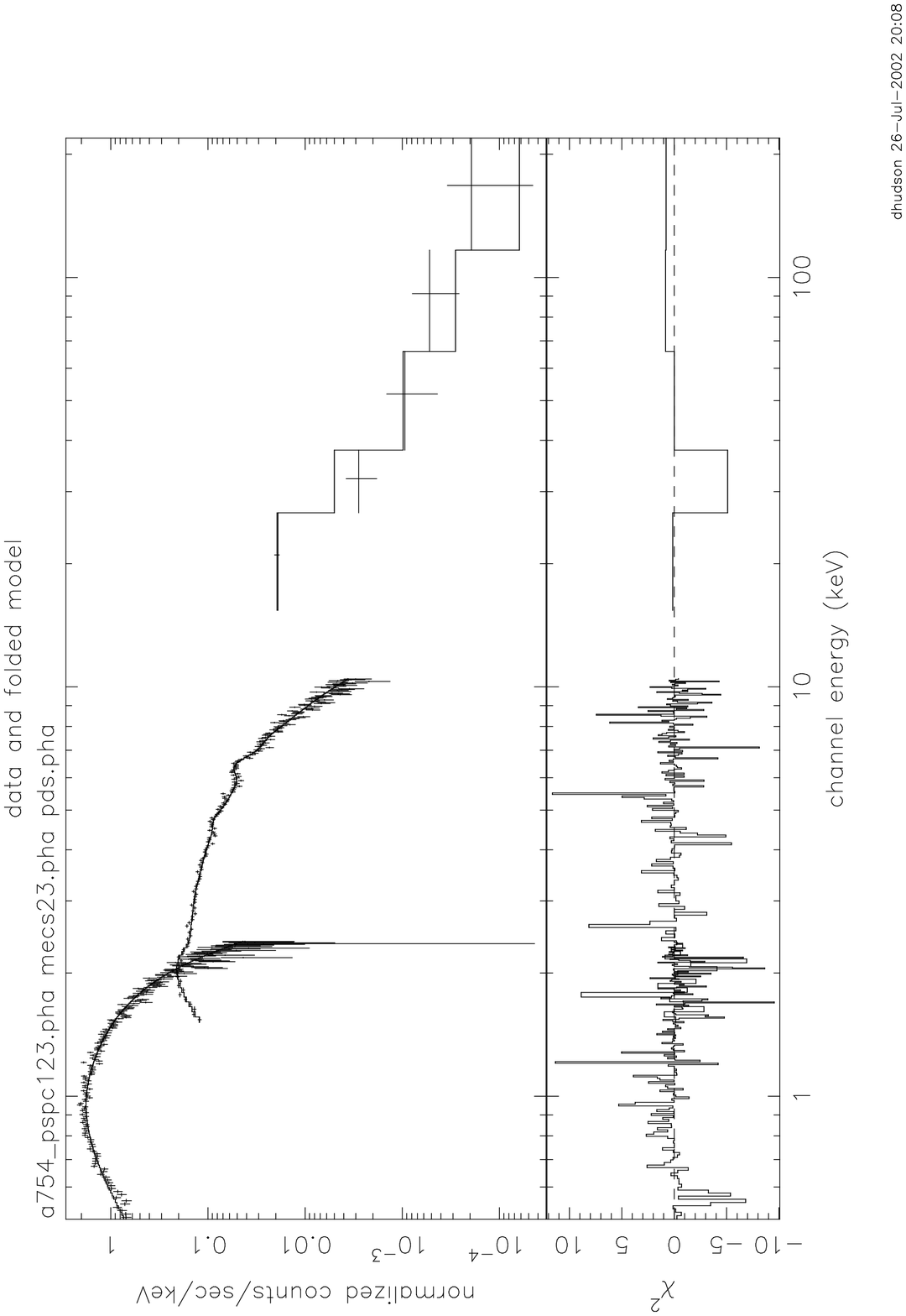}

\plotone{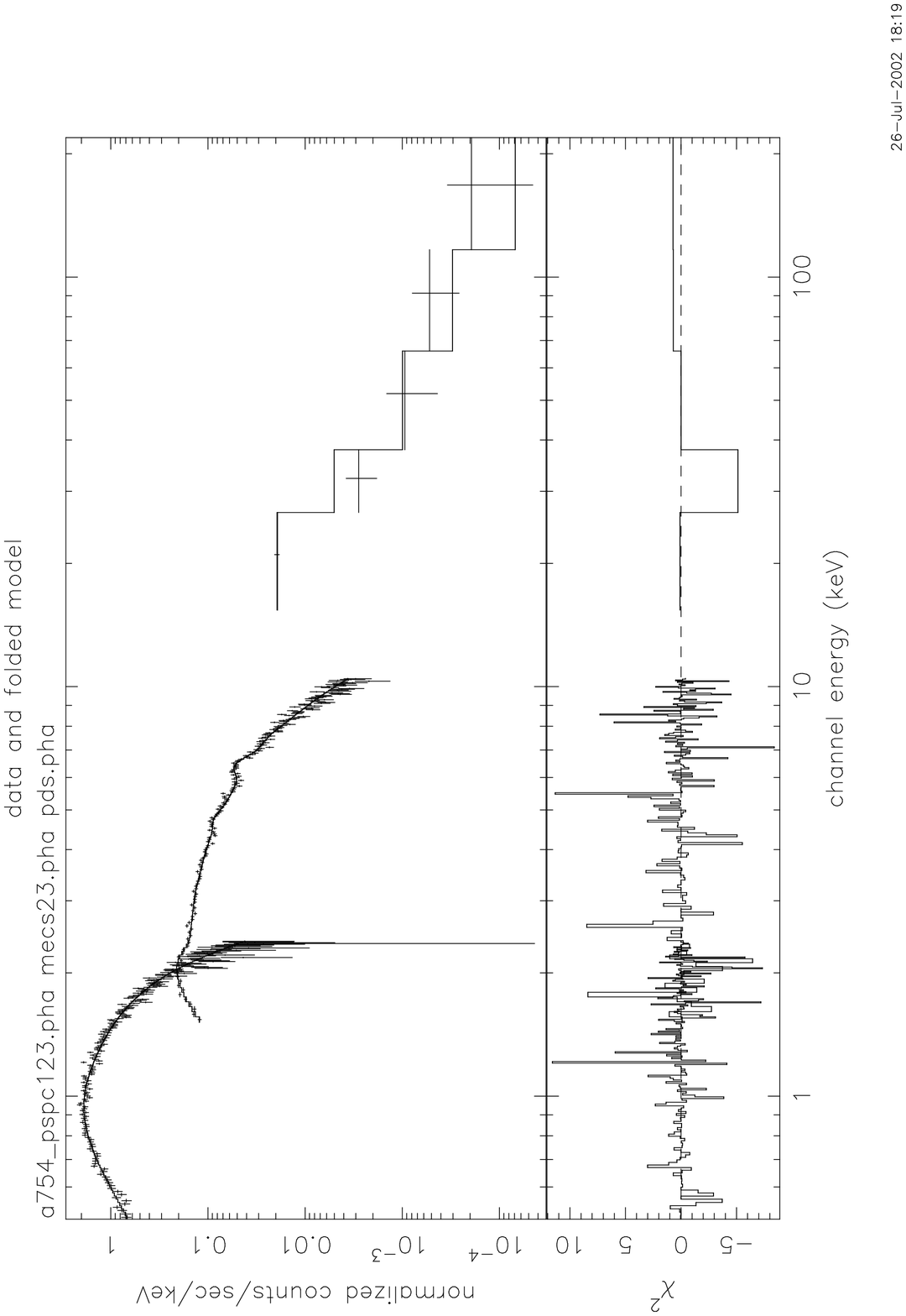}

\plotone{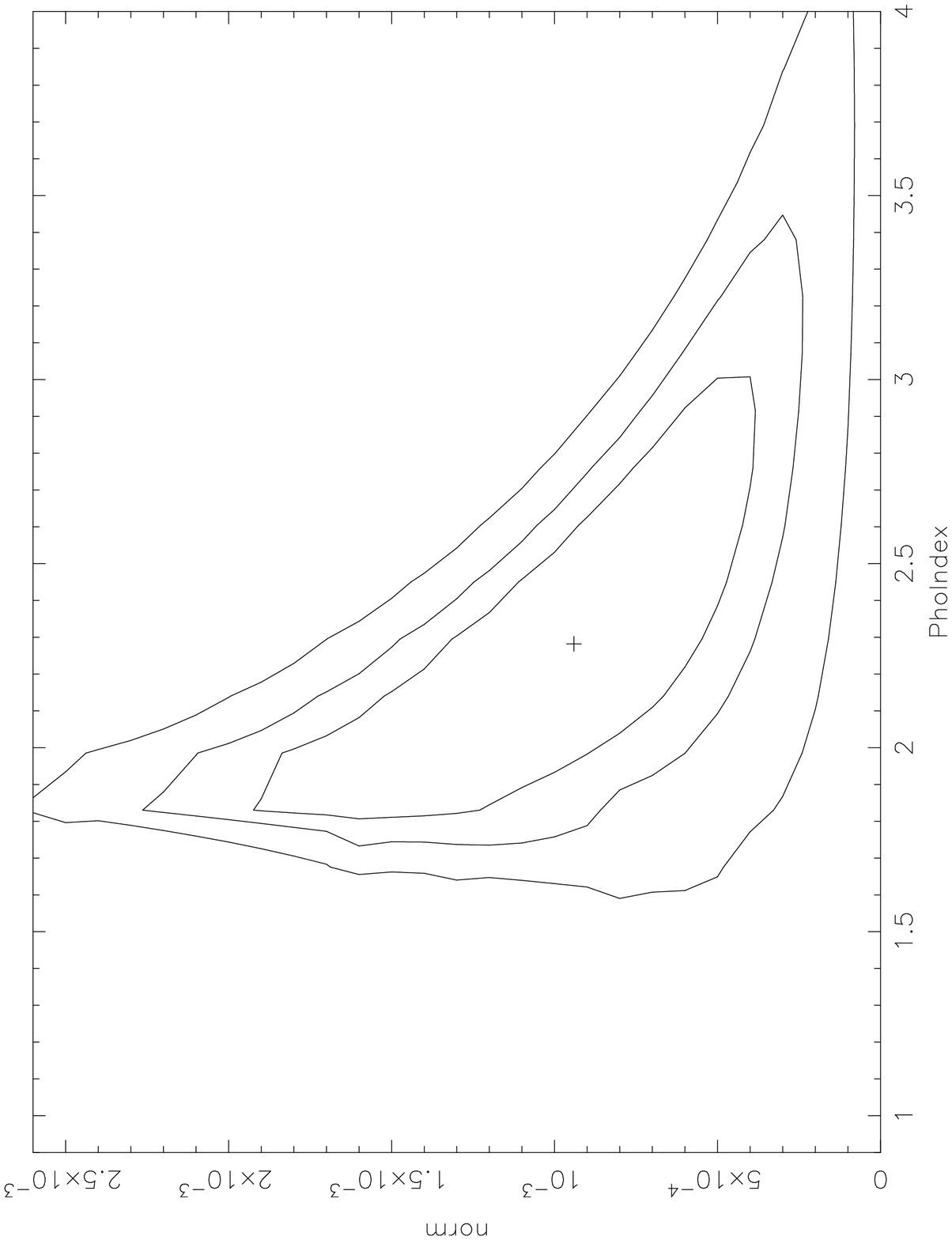}

\plotone{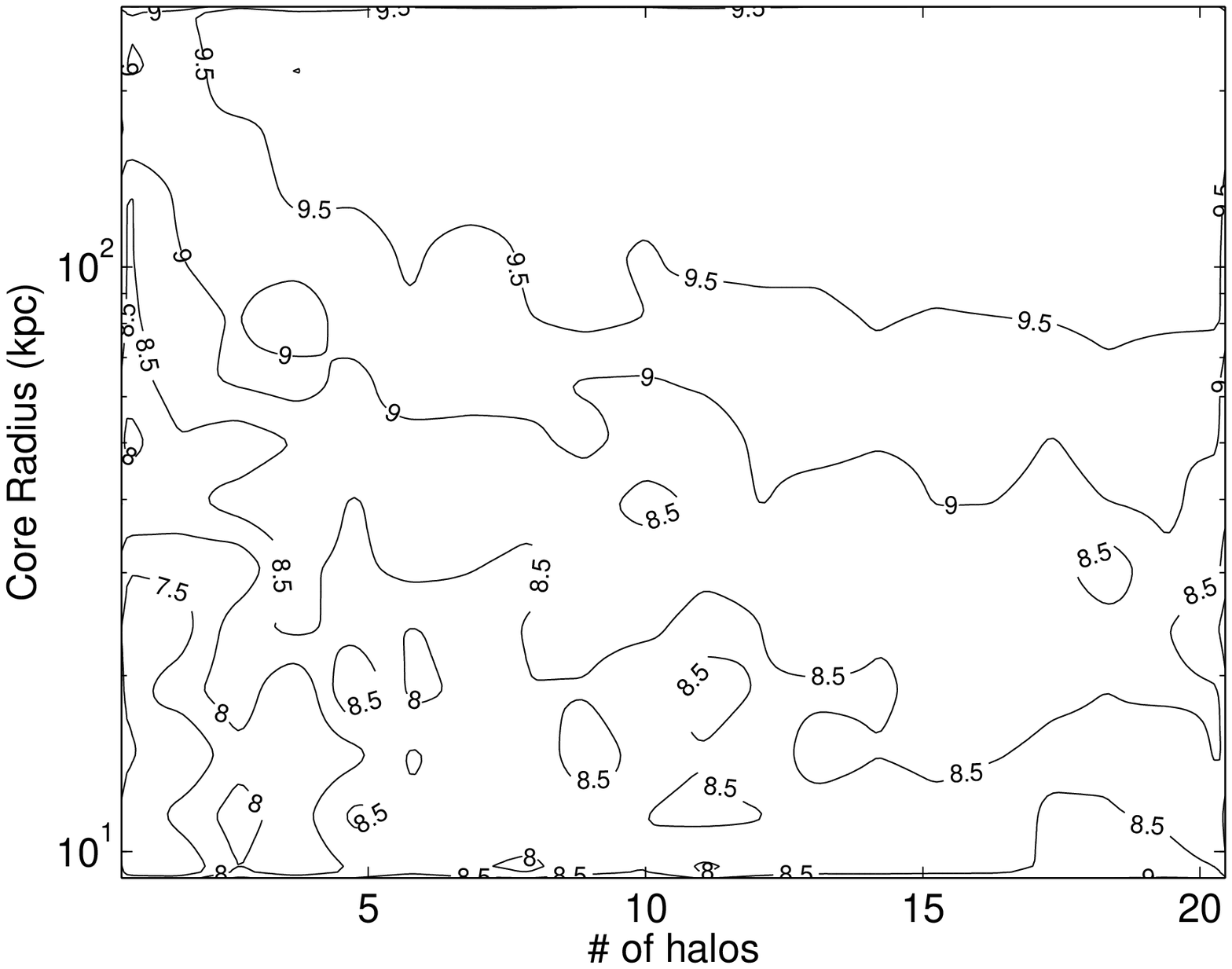}

\plotone{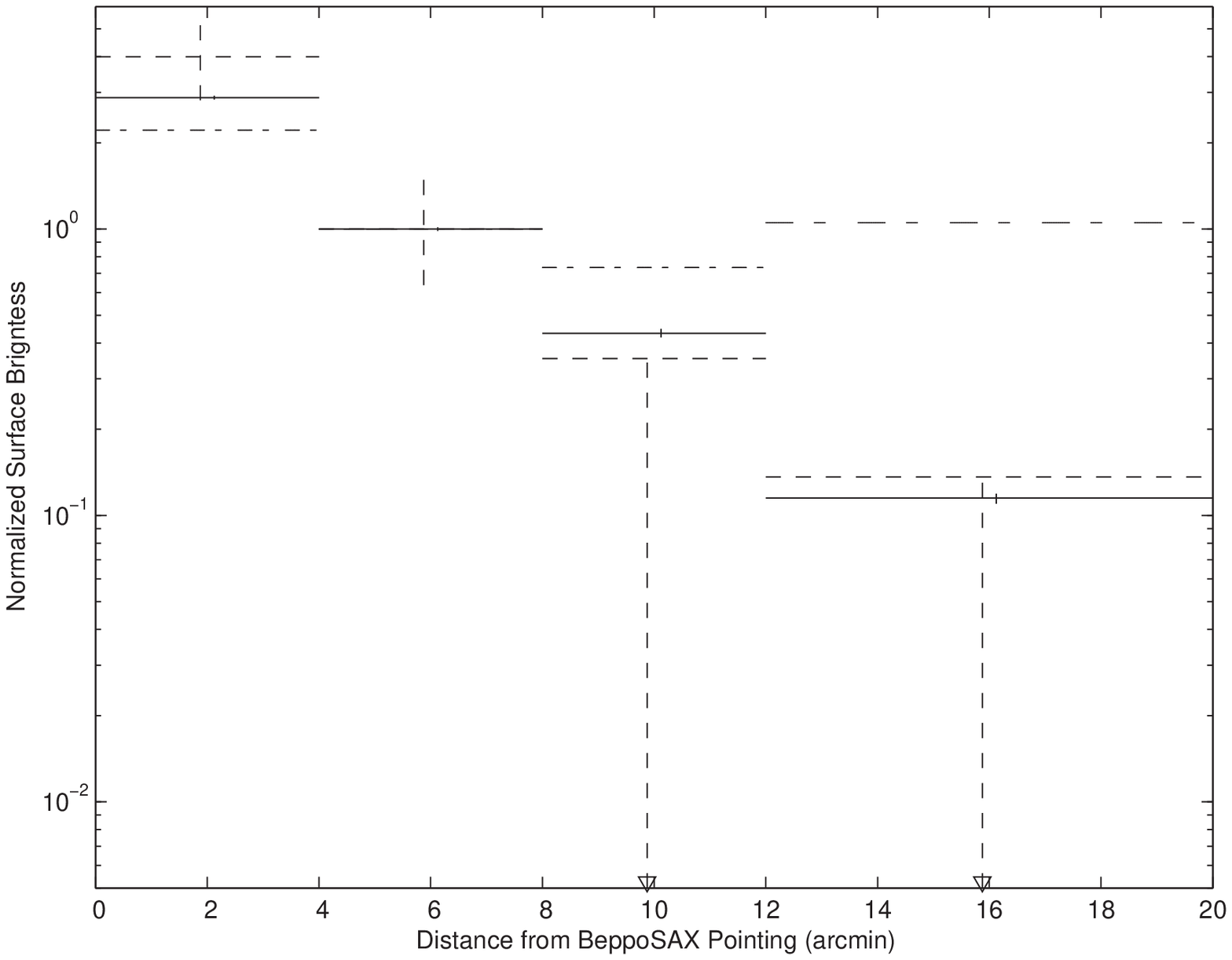}
\end{document}